\documentstyle[12pt,psfig]{article}
\begin{document}
\setcounter{page}{1}
~\\[-2.4cm]
\begin{tabular}{ll}
~\hspace{10.1cm} & {\footnotesize\bf{Fermilab-PUB-96/109-E}} \\
~               & {\footnotesize\bf{KSU-HEP-96-01}}\\
\end{tabular}
~\\[-0.4cm]
\begin{center}
\large
\bf Search for $D^{0}{\overline{D}}^{0}$ Mixing in Semileptonic
Decay Modes
\\[-1.0mm]
\rm
\end{center}
%
\small
%
%
~\\[-9mm]
\begin{center}
    E.~M.~Aitala,$^8$
       S.~Amato,$^1$
    J.~C.~Anjos,$^1$
    J.~A.~Appel,$^5$
       D.~Ashery,$^{14}$
       S.~Banerjee,$^5$
       I.~Bediaga,$^1$
       G.~Blaylock,$^2$
    S.~B.~Bracker,$^{15}$
    P.~R.~Burchat,$^{13}$
    R.~A.~Burnstein,$^6$
       T.~Carter,$^5$
 H.~S.~Carvalho,$^{1}$
  N.~K.~Copty,$^{12}$
       I.~Costa,$^1$
    L.~M.~Cremaldi,$^8$
 C.~Darling,$^{18}$
       K.~Denisenko,$^5$
       A.~Fernandez,$^{11}$
       P.~Gagnon,$^2$
       S.~Gerzon,$^{14}$
       C.~Gobel,$^1$
       K.~Gounder,$^8$
     A.~M.~Halling,$^5$
       G.~Herrera,$^4$
 G.~Hurvits,$^{14}$
       C.~James,$^5$
    P.~A.~Kasper,$^6$
       S.~Kwan,$^5$
    D.~C.~Langs,$^{10}$
       J.~Leslie,$^2$
       B.~Lundberg,$^5$
       A.~Manacero,$^5$
       S.~MayTal-Beck,$^{14}$
       B.~Meadows,$^3$
 J.~R.~T.~de~Mello~Neto,$^1$
    R.~H.~Milburn,$^{16}$
 J.~M.~de~Miranda,$^1$
       A.~Napier,$^{16}$
       A.~Nguyen,$^7$
  A.~B.~d'Oliveira,$^{3,11}$
       K.~O'Shaughnessy,$^2$
    K.~C.~Peng,$^6$
    L.~P.~Perera,$^3$
    M.~V.~Purohit,$^{12}$
       B.~Quinn,$^8$
       S.~Radeztsky,$^{17}$
       A.~Rafatian,$^8$
    N.~W.~Reay,$^7$
    J.~J.~Reidy,$^8$
    A.~C.~dos Reis,$^1$
    H.~A.~Rubin,$^6$
 A.~K.~S.~Santha,$^3$
 A.~F.~S.~Santoro,$^1$
       A.~J.~Schwartz,$^{10}$
       M.~Sheaff,$^{17}$
    R.~A.~Sidwell,$^7$
    A.~J.~Slaughter,$^{18}$
    M.~D.~Sokoloff,$^3$
       N.~R.~Stanton,$^7$
       K.~Stenson,$^{17}$
       K.~Sugano,$^2$
    D.~J.~Summers,$^8$
 S.~Takach,$^{18}$
       K.~Thorne,$^5$
    A.~K.~Tripathi,$^9$
       S.~Watanabe,$^{17}$
 R.~Weiss-Babai,$^{14}$
       J.~Wiener,$^{10}$
       N.~Witchey,$^7$
       E.~Wolin,$^{18}$
       D.~Yi,$^8$
       S. Yoshida,$^{7}$                         
       R.~Zaliznyak,$^{13}$
       and
       C.~Zhang$^7$ \\[1.0mm]

{\normalsize{(Fermilab E791 Collaboration)}} \end{center}

\footnotesize
\it
\begin{center}
$^1$ Centro Brasileiro de Pesquisas F{\'i}sicas, Rio de Janeiro, Brazil\\
$^2$ University of California, Santa Cruz, California 95064\\
$^3$ University of Cincinnati, Cincinnati, Ohio 45221\\
$^4$ CINVESTAV, Mexico\\
$^5$ Fermilab, Batavia, Illinois 60510\\
$^6$ Illinois Institute of Technology, Chicago, Illinois 60616\\
$^7$ Kansas State University, Manhattan, Kansas 66506\\
$^8$ University of Mississippi, University, Mississippi 38677\\
$^9$ The Ohio State University, Columbus, Ohio 43210\\
$^{10}$ Princeton University, Princeton, New Jersey 08544\\
$^{11}$ Universidad Autonoma de Puebla, Mexico\\
$^{12}$ University of South Carolina, Columbia, South Carolina 29208\\
$^{13}$ Stanford University, Stanford, California 94305\\
$^{14}$ Tel Aviv University, Tel Aviv, Israel\\
$^{15}$ 317 Belsize Drive, Toronto, Canada\\
$^{16}$ Tufts University, Medford, Massachusetts 02155\\
$^{17}$ University of Wisconsin, Madison, Wisconsin 53706\\
$^{18}$ Yale University, New Haven, Connecticut 06511\\
[0.3\baselineskip]
\rm
\normalsize
May, 1996
\\[-1.0mm]
\end{center}

\setlength{\baselineskip}{2.6ex}
 
\begin{center}
\parbox{13.0cm}
{\begin{center} {\footnotesize{ABSTRACT\vspace{-2.0mm} }} \end{center}
{\small \hspace*{0.3cm} 
        We report the result of a search for $D^{0}\overline{D} \, ^{0}$
mixing in the 
data from hadroproduction experiment E791 at Fermilab. 
We use the charge of the pion from the strong decay $D^{*+} \rightarrow D^{0}
\pi^{+}$ (and charge conjugate) to identify the charm quantum number of the
neutral $D$ at production, and the charge of the lepton and the
kaon in the semileptonic decays $D^{0} \rightarrow$ $K e \nu$ and $K \mu \nu$
to identify
the charm at the time of decay. 
No evidence of mixing is seen. We set a  90\% confidence level upper 
limit on mixing of $r < 0.50\%$, where
$ r = \Gamma (D^{0}
\rightarrow \overline{D} \, ^{0} \rightarrow K^{+} l^{-} \bar{\nu_{l}})/
\Gamma (D^{0} \rightarrow K^{-} l^{+} \nu_{l})$.
\\[0.7\baselineskip]
\centerline{(Submitted to {\em{Physical Review Letters)}} }
}}
\end{center}
\normalsize

        The predicted rate of $D^{0} \overline{D} \, ^{0}$ mixing in the
Standard Model 
 \cite{mix-standard} is several
orders of magnitude below the sensitivity of current experiments. 
However, several theoretical extensions \cite{beyond-sm} to the
Standard  Model predict $D^{0} \overline{D} \, ^{0}$   mixing rates 
high enough to be measurable by existing experiments, making it
interesting to search for this process.
The mixing rate is parametrized as 
$ r = \Gamma (D^{0}
\rightarrow \overline{D} \, ^{0} \rightarrow \bar f)/\Gamma (D^{0}
\rightarrow f)$, where $f$ is the final decay state 
used to identify the charm quantum number of the neutral $D$ at the
time of decay. We report here a limit on $r$ using semileptonic decays     
in the data from Fermilab experiment E791. 

Many experiments have used hadronic $D^{0}$ decays to search for mixing. 
For example, Fermilab experiment E691
studied $D^{0} \overline{D} \, ^{0}$ mixing by looking for the decay
chain $D^{*+}$ $\rightarrow \pi^{+}D^{0}$, followed by 
$ D^{0}\rightarrow \overline{D} \, ^{0} \rightarrow K^{+}\pi^{-}$ or
$K^{+}\pi^{-} \pi^{+} \pi^{-}$ \cite{mixe691}. 
A wrong-sign
charged $K$ from the neutral $D$ decay ({\em e.g.,} $D^{0} \rightarrow
\overline{D} \, ^{0} \rightarrow K^{+} \pi^{-}$) can be a signature of
mixing. However, a
wrong-sign $K$ can also come from doubly-Cabibbo-suppressed
(DCS) decays in which a $D^{0}$ decays directly into the wrong-sign
kaon ({\em e.g.}, $D^{0} \rightarrow K^{+} \pi^{-}$). Moreover, the DCS
amplitude can interfere with the mixing
amplitude, reducing the sensitivity to mixing
 \cite{guymix} even though the mixing, DCS, and
interference terms in 
principle can be
separated statistically using decay-time information. 
E691 reported a 90\% confidence level (C.L.) limit on $r$ of 0.37\%
assuming no  
interference between DCS and mixing amplitudes.
For  worst-case interference, their limit 
is $1.9\%$ in the $K\pi$ mode and $0.7\%$ in the $K\pi\pi\pi$  
mode. 
E791 has used the same decay modes to study $D^{0}\overline{D} \, ^{0}$ mixing
using a much larger data sample and more general assumptions \cite{e791mix}.
CLEO II has observed a wrong-sign signal in the mode
$D \rightarrow K\pi$, and measures the ratio of the wrong-sign to
right-sign decays to be ($0.77 \pm 0.25 \pm 0.25$)\% \cite{mix:cleo}. 
However, an unambiguous mixing signal cannot  be established 
from the CLEO result because of the lack of decay-time 
information. 

An alternate way to make a mixing measurement is to  use semileptonic
decays to tag the charm  of the neutral $D$ at the time
of decay. There is no DCS amplitude in these decays,  eliminating
the complications of interference. 
Fermilab experiment E615 
searched for mixing by
looking for pairs of muons with the same charge in a 
single event \cite{mixe615}. Same-sign 
muons could come from the semileptonic decays of a $D$ meson ($D^{0}$
or $D^{+}$) and a $\overline{D} \, ^{0}$ that has oscillated into a $D^{0}$. 
E615 obtained a 90\% C.L.
upper limit on $r$ of 0.56\% using specific assumptions for charm
production cross sections and $D$ branching fractions.  

In this letter we report the result of a mixing search using   
reconstructed semileptonic decays of the $D^{0}$ in the data sample of 
hadroproduction experiment E791 at Fermilab. We observe a large signal
for the right-sign (RS) decay chain 
$D^{*+} \rightarrow \pi^{+} D^{0} \rightarrow
\pi^{+}(K^{-} l^{+} \nu_{l})$, where $l$ is an $e$ or a $\mu$, in
which the charge of the  $\pi$ is 
the same as the charge of the lepton from the neutral D decay 
(charge conjugate modes are implied throughout this paper). We
search for mixing in wrong-sign (WS) $D^{*+}$ decay candidates
in which the charge of the $\pi$ is opposite to that of the
charged lepton.
These candidates  could correspond to the decay chain $D^{*+} \rightarrow
\pi^{+}D^{0} \rightarrow \pi^{+} \overline{D} \, ^{0} \rightarrow
\pi^{+}(K^{+} l^{-} \bar{\nu_{l}})$. We look for two signatures
of mixing in the WS sample -- a peak in $Q$-value 
$ (Q \equiv M(Kl\nu\pi) - M(D^{0}) - M(\pi))$ at about 5.8 MeV
and the characteristic distribution in proper decay time $t$. Assuming
mixing is small, the time evolution of 
the mixed $D$'s is given by $dN/dt \propto t^{2}e^{-\Gamma t}$, where
$\Gamma$ is the $D^{0}$ decay rate.

The E791 experiment \cite{e791-det}
recorded $2 \times 10^{10}$ events 
from 500 GeV/$c$ $\pi^{-}$ interactions in five
thin targets (one platinum, four diamond) separated by gaps of 1.34 to
1.39 cm. 
Precision vertex and tracking information was
provided by 23 silicon microstrip detectors and 35 drift chamber
planes. Momentum
was measured using two dipole magnets. 
Two segmented threshold Cerenkov counters provided $\pi/K/p$ separation
in the 6 - 60 GeV/$c$ momentum range \cite{cerenkove}. 

A segmented lead and liquid-scintillator calorimeter 
is used to identify 
electrons from their energy deposition and transverse shower shape. 
For the cuts used in this analysis, the
probability that a $\pi$ is misidentified as an electron is typically 1.6\%
and the probability that a $K$ is misidentified as an
electron is typically 0.8\%. 
Muon identification is provided by
two planes of scintillation counters, located behind shower-absorbing
calorimeters and steel
shielding with a total thickness equivalent to 2.5 meters of steel (15
proton interaction lengths). All muon candidates are required to have 
momentum greater that 10 GeV/$c$ to reduce background from decays in
flight, and  to 
leave a signal in the expected scintillation counters, allowing for
multiple scattering.
For the cuts used in this analysis, the
probability that a $\pi$ is misidentified as a $\mu$ is typically 3.6\% and
the probability that a $K$ is misidentified as a $\mu$ 
is typically 4.6\%.

To reduce background, a candidate $D^{0}$ decay vertex is required to be 
separated from the production vertex by at least $8\sigma_{z}$, where
$\sigma_{z}$ 
is the error on the separation between the two vertices. 
The decay vertex is required to be at least  $3\sigma$ away from the 
edge of the nearest solid material, where $\sigma$ is the error on the
separation. 
The minimum parent
mass, defined as $M_{min} = p_{T} + \sqrt{p_{T}^{2} + M_{Kl}^{2}}$, 
where $p_{T}$ is the transverse momentum of the $Kl$ with respect to
the direction of flight of the $D^{0}$ and $M_{Kl}$ is the invariant
mass of the $Kl$ candidates \cite{m-min}, is required to be in the range
1.6 to 2.1 GeV/$c^{2}$.
The $M_{min}$ distribution for Monte Carlo signal events has a cusp at
the $D^{0}$ 
mass, and falls   
rapidly at lower values of $M_{min}$, whereas background rises as 
$M_{min}$ decreases. We also require the 
invariant mass of the $Kl$ candidate, $M_{Kl}$, to be in the range 1.15 to
1.80 GeV/$c^{2}$. The lower cut on $M_{Kl}$ reduces noncharm
background, and the
upper cut on $M_{Kl}$ removes feedthrough from  $D^{0} \rightarrow
K\pi$ decays
into the RS sample,
in which the $\pi$ is misidentified as a lepton. 
We require the transverse momentum of the lepton
with respect to the direction of flight of the candidate $D^{0}$ to be
greater than 
0.2 GeV/$c$ and that of the hadron to be greater than 0.4 GeV/$c$,
since charm decay products tend to have larger such transverse momenta
than background tracks. 
The $\pi^{+}$ track
from the $D^{*+}$ is required to be consistent with belonging to 
the primary vertex and to  
have momentum greater than 2 GeV/$c$.  

To eliminate feedthrough of the $K\pi$ mode into the
wrong-sign sample through double misidentification of the hadrons 
(the $K$ 
misidentified as a lepton and the $\pi$ misidentified as a $K$),
we require $|M_{\pi K} - M_{D^{0}}| > 30$ MeV/$c^{2}$ (typical $\sigma$
of the $D^{0}$ mass peak in the $K\pi$ mode is 15 MeV/$c^{2}$), where
$M_{\pi K}$ is the invariant mass of the $Kl$ candidate when the $K$
is assigned the $\pi$ mass and the $l$ is assigned the $K$ mass. 

Additional cuts are applied to $K \mu \nu$ candidates 
because kaons and pions are more likely to be misidentified as muons
than as electrons due to punchthrough and decays in flight.
If the muon track is positively identified as a kaon in the Cerenkov
detectors, the decay vertex is rejected.
Feedthrough from the mode $D^{0}
\rightarrow K^{-} K^{+}$ is eliminated by the requirement
$|M_{KK}-M_{D^{0}}| > 30$ MeV/$c^{2}$, where $M_{KK}$ is the invariant mass
of the $Kl$ candidate when both tracks are assigned the $K$ mass. We 
also demand that there be
one and only one $D^{*}$ candidate ($Q$-value $<$ 80 MeV/$c^{2}$) in
each event in the $K \mu \nu$ sample.    
An event is
rejected if more than one $D^{*}$ candidate is found.

\begin{figure}[htbp]
\centerline{\psfig{figure=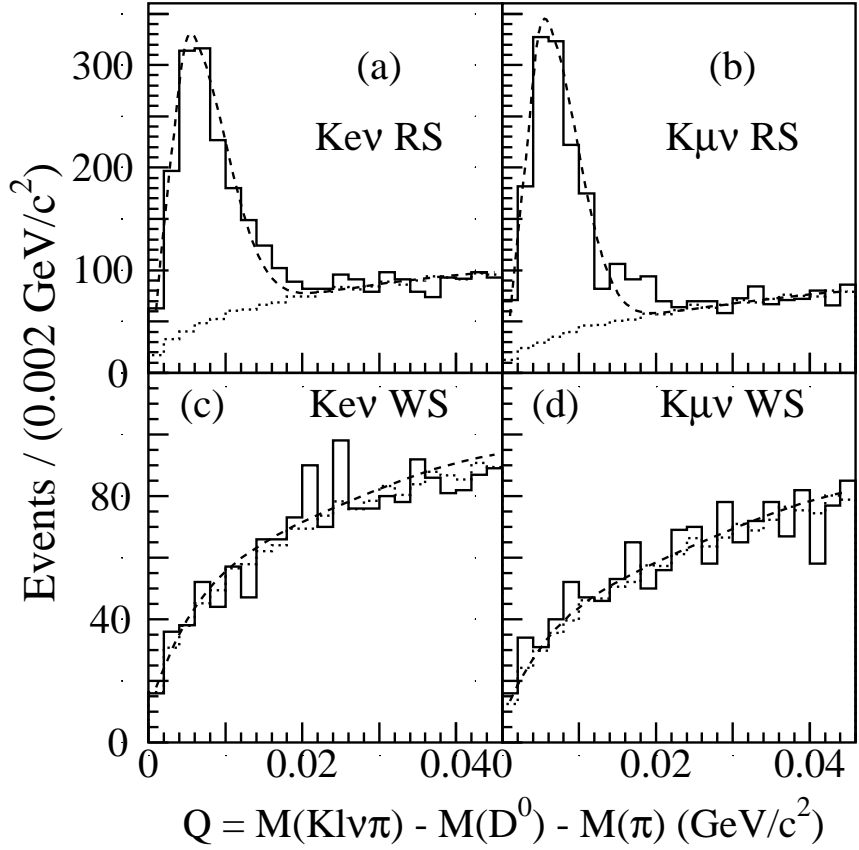,width=4.5in}}
        \caption{ The $Q$-value distributions for (a) $K e \nu$ RS, (b)
$K \mu \nu$ RS, (c) $K e \nu$ WS, and (d) $K \mu \nu$ WS candidates. 
The solid line histograms show the data $Q$-value distributions, the
dashed lines are the projections of the fit in $Q$-value, and the  
dotted lines show the $Q$-value distribution obtained from
combining a $D^{0}$ from one event and $\pi$ from another, normalized to
the number of events with $Q >$ 0.025 
GeV/$c^{2}$ in the respective histograms. }
\label{qval}
\end{figure}

        Since there is an undetected neutrino in a semileptonic decay,
the  $D^{0}$ momentum cannot be reconstructed directly. 
However, using the measured positions of the primary and secondary
vertices, the measured $K$ and $l$ momenta, 
and assuming the parent 
particle mass is that of a $D^{0}$, one can solve for the neutrino 
momentum up to a two-fold ambiguity. 
The solution resulting in higher $D^{0}$ momentum is used for all events.
Monte Carlo (MC) 
studies indicate that it gives a better estimate of the true
momentum for the selected sample.
From MC, we determine that the root mean square (r.m.s.)  deviation
between the calculated
and the true $D^{0}$ momenta is about 15\%. This also causes smearing
in the calculated proper decay time. The effect of this smearing is 
discussed below. 
Having obtained the $D^{0}$ momentum,
we calculate the invariant mass of the $D^{*+}$  candidate and the proper
decay time of  
the $D^{0}$ candidate.
The final $Q$-value distributions for $K e \nu$ and $K \mu \nu$ candidates 
are shown in Figure~\ref{qval}. 

To search for mixing signals,
separate unbinned maximum likelihood fits are performed on the $K e \nu$ and
$K \mu \nu$ samples,  using the
$Q$-value and proper decay time $t$ for each event.
The expected $Q$-value signal shape in WS data is obtained directly from 
fits to the large, kinematically-identical RS signal.
It is parametrized by asymmetric Gaussian distributions, broader on
the high $Q$ side, 
with widths $\sigma$ decreasing with longer proper decay time $t$,
$\sigma(t) = \sqrt{(\sigma_{0})^{2} + (C/t)^{2}}$. This $t$-dependence arises
because the $D^{0}$ direction is measured better for decays at greater
distances. 
The $Q$-value distribution of the background under the $D^{*}$ peak is
described  by the spectrum which results from combining  a $D^{0}$
candidate from one event with pions from other events   
to form random $D^{0}-\pi$ mass combinations (dotted histograms in
Figure~\ref{qval}).

\begin{figure}[htbp]
\centerline{\psfig{figure=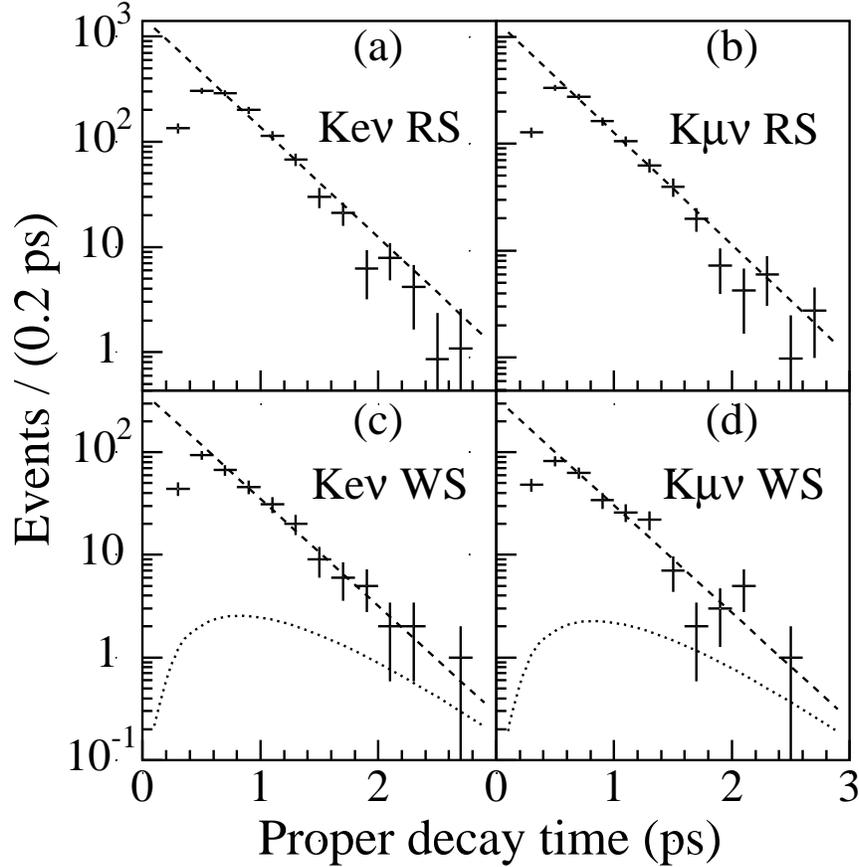,width=4.5in}}
        \caption{Dependence of RS and WS signals on proper decay time $t$.
Crosses represent the measured decay-time distributions for (a) $K e \nu$
and (b) $K \mu \nu$ background-subtracted RS signals, and for (c) $K e \nu$ 
and (d) $K \mu \nu$ WS signal region ($Q < 0.015$ GeV/$c^{2}$).
The dashed line in
all histograms is the expected $D^{0}$ decay-time distribution
uncorrected for 
detector acceptance, normalized to the number of events with $t > 0.7$
ps (where acceptance is uniform). The dotted
lines in (c) and (d) represent the 
expected decay-time distributions uncorrected for detector acceptance
 for $K e \nu$ and $K \mu \nu$ mixing
signals,
corresponding to our 90\% C.L. limit in each mode. } 
\label{tdec}  
\end{figure}

When smearing is neglected, the measured proper decay-time spectrum of a mixing
signal is proportional 
to $t^{2} \epsilon(t)e^{-\Gamma t}$, where $\epsilon(t)$ is the $t$-dependent
detector efficiency. This spectrum is obtained by multiplying 
the measured distribution
(crosses in Figure~\ref{tdec} (a) and (b)) of background-subtracted RS
data ($\propto 
\epsilon(t)e^{-\Gamma t}$) by the 
mean value of $t^{2}$ in each bin. The $t$-distribution of the non-$D^{*}$
background is obtained from data events in the $Q$-value sideband with $25 < Q
< 60$ MeV$/c^2$. Distributions of $t$ in the three $Q$-value sidebands
$20 < Q < 30$, $40 < Q < 50$, and $ 60 < Q < 70$ MeV$/c^2$ are 
 identical within statistical errors. Therefore, the background
decay-time distribution is constant across the $Q$-value spectrum,
as expected if most  background is due to real $D^{0}$'s combined with 
random pions. 
Sideband $t$ dependence thus can be and is used to model the background in the
signal region.

From the fits, we find $N_{RS}=1237\pm 45$ RS events and
$N_{mix}=4.4^{+11.8}_{-10.5}$ WS 
mixed events in the $K e \nu$ samples, and $N_{RS}=1267\pm 44$ RS events and
$N_{mix}=1.8^{+12.1}_{-11.0}$ WS mixed events in the $K \mu \nu$ samples. 
There is
no indication of a mixing signal.

The mixing rate is $r = (N_{mix}/N_{RS}) \times \alpha$, where
$\alpha$ accounts for 
the dependence of detector acceptance on $t$ and
the different $t$-dependences of mixed and unmixed decays. Since
vertex reconstruction 
efficiency is low at small $t$, the detector is more efficient at finding
the longer-lived mixed decays.
Specifically,
$\alpha \equiv  (\Gamma \int_{0}^{\infty}  \epsilon (t) e^{-\Gamma t}
dt)\, / \, (\frac{1}{2} \Gamma^{3} \int_{0}^{\infty} \epsilon (t)
t^{2} e^{-\Gamma t} dt)$. 
It is measured from  the background-subtracted RS decay-time
distribution ($\propto \epsilon(t)e^{-\Gamma t}$) using numerical integration.
Values of $\alpha$ are $0.44 \pm 0.02$ for the $K e \nu$ mode and
$0.46 \pm 0.02$ for the $K \mu \nu$ mode. 

We measure the mixing rate to be  $r =
(0.16^{+0.42}_{-0.37})\%$ for the $K e \nu$ mode and $r =
(0.06^{+0.44}_{-0.40})\%$ 
for the $K \mu \nu$ mode. Taking the weighted average of these two
statistically-independent results, we get an average mixing rate of
$r= (0.11^{+0.30}_{-0.27})\%$.
This gives an upper limit for $D^{0}\overline{D} \, ^{0}$ mixing of $r
< 0.50\%$ at the 90\% confidence level.

Since right- and wrong-sign data samples are selected using
identical criteria, most systematic uncertainties
cancel in  the mixing rate.
Two possibly significant sources of systematic error, the time
resolution for a mixing signal and feedthrough of hadronic decays, remain.

The decay-time distribution used in the fit for mixed events
is proportional to $\epsilon(t) t^{2} e^{-\Gamma
t}$, which is valid  only if the decay times are measured exactly. 
Due to finite resolution of the detector and the
choice of 
one of the two neutrino momentum solutions, the measured mixed
decay-time distribution differs slightly from the distribution used in
the fit. The r.m.s. deviation of the
measured $t$ from the true one is about 15\%. 
MC studies indicate that the effect of this smearing in proper decay
time on the final result is less than 10\% of the statistical error
and hence is ignored.

Feedthrough of hadronic decays into the semileptonic sample is expected to come
mainly from modes such as
$D^{0} \rightarrow K^{-} \pi^{+} \pi^{0}$ in which an undetected neutral
hadron approximates the kinematics of a missing neutrino, and a hadron is
misidentified as a lepton.
Misidentified hadronic decays have two possible effects: 
1) feedthrough into the WS signal would give a false mixing signal;
2) feedthrough into the RS signal
would inflate the denominator of the mixing rate, thus overestimating our
sensitivity to mixing. 
WS feedthrough requires $K$-$\pi$ misidentification as well as hadron-lepton
misidentification. Another source of feedthrough into the WS sample
is doubly misidentified semileptonic decays (in which the hadron is
misidentified as a lepton and the lepton is misidentified as a kaon). 
We see  no WS signal and  make no correction for
this effect.
RS feedthrough has been modeled by Monte Carlo simulation, and shown to
be about 3\% of the RS signal. No correction is made for this effect.

To check our fitting procedure and to look for possible systematic effects in
our sensitivity, we added  
fixed numbers $M$ ($M$ = 10, 20, 30, 40; the typical fit error on
our $N_{mix}$ is 10) of simulated mixed  events to the wrong-sign data sample
and refit. These fits found a mixing
rate systematically (10 - 15)\% higher than the correct value, mainly due
to an overestimate of the correction factor $\alpha$ which is a 
result of our choice of the neutrino momentum resulting in higher
$D^{0}$ momentum. We
conservatively choose not to
correct for this systematic overestimate of mixing rate. 

In summary, we have searched for $D^{0}\overline{D} \, ^{0}$ mixing using
$D^{*+}$ $\rightarrow \pi^{+}D^{0} \rightarrow \pi^{+} \overline{D} \, ^{0}
\rightarrow \pi^{+}(K^{+}l^{-}\bar{\nu_{l}})$ 
candidates together with decay
time information. We obtain a 90\% C.L.
upper limit
of 0.50\% on the mixing rate. 
This result is the best model-independent limit on
$D^{0}\overline{D} \, ^{0}$ mixing to date. 


We gratefully acknowledge the assistance of the staff of Fermilab and of all
the participating institutions.  This research was supported by the Brazilian
Conselho Nacional de Desenvolvimento Cient\'{i}fico e Technol\'{o}gico,
CONACyT (Mexico), the Israeli Academy of Sciences and Humanities, 
the U.S. Department of Energy, the U.S.-Israel
Binational Science Foundation, and the U.S. National Science Foundation.
Fermilab is operated by the Universities Research Association, Inc., under
contract with the United States Department of Energy.

\small
\bibliographystyle{unsrt}

\end {document}